\documentstyle[12pt]{article}
\newcommand{\n}{\noindent}

\begin{document}
\baselineskip 0.657 true cm

\begin{titlepage}

\begin{flushright}

PUPT-1321

\end{flushright}

\begin{flushright}

May 1992

\end{flushright}
\begin{center}

\vspace{.5 true  cm}

\Large
{\bf The Critical Point of Unoriented Random }\\
\vspace{6pt}
{\bf Surfaces with a Non-Even Potential}\\

\vspace{3 true  cm}

\large
{M.A. Mart\'{\i}n-Delgado }\\

\vspace{0.5 true  cm}

\normalsize
{\em  Joseph Henry Laboratories } \\
{\em Princeton University}\\
{\em Princeton, NJ 08544}

\vspace{.5 true  cm}

\end{center}

\begin{abstract}
The discrete model of the real symmetric one-matrix ensemble is
analyzed with a cubic interaction. The partition function is
found to satisfy a recursion relation that solves the model.
The double scaling-limit of the recursion relation leads to a
Miura transformation relating the contributions to the free
energy coming from oriented and unoriented random surfaces.
This transformation is the same kind as  found with a
cuartic interaction.
\end{abstract}

\vspace{5 true cm}
\hrule
\vskip.2cm
\noindent

\noindent

\hbox{ $\star $ Email: martind@puhep1.princeton.edu}
\hbox{ Research supported in part by SP B.F.P.I.E. fellowship. }

\eject

\newpage

\end{titlepage}

{\bf 1. Introduction}

Matrix models have been used to study the formulation of non-perturba\-tive
two-dimensional Euclidean quantum gravity (pure or coupled with conformal
matter with $c\leq 1$) following the seminal work of references \cite{B-K},
\cite{D-S}, \cite{G-M}. These models may also be viewed  as models of random
surfaces
or zero-dimensional string theory. The partition function of the Hermitian
matrix ensemble is identified as the theory of oriented surfaces. This
connection is
established by identifying the dual to the Feynman diagrams of the
matrix model with random discretizations of two-dimensional surfaces of
arbitrary genus. The technique of the double-scaling limit allows us to recover
the
continuum limit of the theory \cite{B-K}, \cite{D-S}, \cite{G-M}.

Many other one-matrix models have been solved following the tools developed
for the Hermitian ensemble \cite{Per-S}, \cite{My-P1}, \cite{My-P2}. In this
paper we shall consider the real symmetric one-matrix model \cite{Bre-N},
\cite{Bre-N2},
\cite{Har-M}, \cite{Meh-M} introduced to describe theories with unoriented
random surfaces. In these treatments, the potential in the matrix model was for
simplicity
taken to be even. When this is the case, the orthogonal polynomials suitable to
 solve the model happen to have a definite parity, thereby providing a
computational simplification similar to the one appearing in the standard
Hermitian
model \cite{Bac-P}. In this letter we shall address the problem of a non-even
 potential which we solve for the discrete model with a cubic interaction in
sections  2,3,4.

This study is motivated by the equivalent problem taking place in the Hermitian
ensemble where it is found \cite{Sim}, \cite{Vis} that the continuum limit
of the model with a cubic potential leads to a string susceptibility satisfying
the Painlev\`e I equation, as it happens with a cuartic potential. In section
 5 the double-scaling limit of a real symmetric matrix ensemble with a cubic
term is computed and compared with the cuartic potential case of \cite{Bre-N},
\cite{Har-M}.

 Our starting point is the partition function for the real symmetric
$2N\times 2N$ matrices $M^T=M$,

\begin{equation}
{\cal Z}_{2N} = e^{{\cal F}_{2N}} = \int \prod _{0\leq i\leq j\leq 2N-1}
[dM_i^j]\ e^{-\frac{\beta }{2}tr(V(M))}            \label{1}
\end{equation}

\n with the matrix potential given by

\begin{equation}
V(M) = \frac{g_2}{2}\  M^2 + \frac{g_3}{3}\ M^3      \label{2}
\end{equation}

The identification of (\ref{1}) as a theory of unoriented random surfaces comes
from the perturbation expansion of the interaction term in (\ref{1}), (\ref{2})
around the gaussian ensemble $g_3=0$. The propagator is

\begin{equation}
<M_i^j M_m^n>_0 = g_2^{-1} (\delta _{im}\delta ^{jn} +
\delta _i^n\delta _m^j)                              \label{3}
\end{equation}

The first term in (\ref{3}) represents the ``usual" propagator in the double
line notation of the Feynman diagrams, while the second one represents a
``twisted" propagator. The latter contribution amounts to the loss of
orientability of the triangulated surface associated to the dual of a
Feynman graph\footnote{To be more precise, in the continuum limit it is also
needed that both contributions in (\ref{3}) have the same positive weight
 \cite{Bre-N}.}. Each connected vacuum diagram is weighed by a factor of
$N^{\chi }$, where $\chi  $ is the Euler characteristic of the surface, which
now can take both even and odd values. Thus, all kinds of non-orientable
surfaces contribute to the sum of the random surfaces described by
(\ref{3}) \cite{Bre-N}, \cite{Har-M},

\begin{equation}
{\cal Z}_{2N} = N^2 {\cal Z}_{sphere} + N^1 {\cal Z}_{RP^2} +
N^0 ({\cal Z}_{torus} + {\cal Z}_{Klein\ bottle}) + \cdots  \label{4}
\end{equation}

{\bf 2. The Method of Skew Orthogonal Polynomials}

The parttion funtion of non-orientable pure gravity is computed with the help
of an appropiate set of polynomials in a similar way to what happens in the
case of the Hermitian matrix model. Thus, when we integrate over ``angles",
i.e., over the real orthogonal group with the action $M=OXO^{-1}=OXO^{T}$ in
(\ref{1}), it is possible to express the partition function in terms of the
eigenvalues $X=diag(x_1,\ldots ,x_{2n})$ of the matrix $M$. The result is
\cite{Meh1}

\begin{equation}
{\cal Z}_{2N} = K \times \int \prod _{i=1}^{2N}dx_i |\Delta(x)|
e^{-\frac{\beta }{2}\sum _{i=1}^{2N} V(x_i)}                 \label{5}
\end{equation}

\n where $V(x)$ is given by (\ref{2}). The main new feature is the appearence
of the Vandermonde determinant $\Delta (x)= \prod _{i<j} (x_i-x_j)$ in
absolute value and raised to the power of one, rather than the usual factor
$\Delta ^2 (x)$ showing up in the Hermitian matrix model. This is a major
difference between both models as far as the computation of (\ref{5}) is
concerned. From the symmetry properties of (\ref{5}) we may write
\cite{Meh1}

\begin{equation}
{\cal Z}_{2N} = K (2N)! \int _{\stackrel{\cdots }{R(-\infty,x_1,\ldots
,x_{2N},\infty)}}
 \int \prod _{i=1}^{2N}
e^{-\frac{\beta }{2}\sum _{i=1}^{2N} V(x_i)}\times
[det (x_j^{i-1})]_{i,j=1,\ldots ,2N}                         \label{6}
\end{equation}

\n here the region of integration $R$ is $-\infty <x_1\leq \ldots \leq
x_{2N}<\infty $. Note that the absolute value in (\ref{6}) is no longer
required. As usual,
 it is convenient to introduce monic polynomials $R_i(x)$ in order to rewrite
the Vandermonde determinant as

\begin{equation}
\Delta (x) =
[det (x_j^{i-1})]_{i,j=1,\ldots ,2N} =
[det R_{i-1}(x_j)]_{i,j=1,\ldots ,2N}                          \label{7}
\end{equation}

We want to solve the partition function ${\cal Z}_{2N}$ in terms of quantities
related to these monic polynomials. To do that, we have to chose the $R_i(x)$
as skew orthogonal polynomials with respect to the following anti-symmetric
scalar product \cite{Meh-M}

\[ <R_i,R_j>_R \equiv \int _{-\infty }^{\infty } \int _{-\infty}^{\infty}
 dx dy \epsilon (x-y)
e^{-\frac{\beta }{2}[V(x)+V(y)]}\]
\begin{equation}
\equiv r_{[i/2]}\ z_{ij}                                      \label{8}
\end{equation}

\n where the only non-zero values of $z_{ij}$ are

\begin{equation}
z_{2i,2i+1} = -z_{2i+1,2i} = 1                              \label{9}
\end{equation}

\n and

\begin{equation}
\epsilon (y-x) = \left\{ \begin{array}{cc}
                 \frac{1}{2} & \mbox{if $y>x$} \\
                \frac{-1}{2} & \mbox{if $y<x$}
                  \end{array}
                 \right.                                    \label{10}
\end{equation}

Using Mehta's method \cite{Meh1} of integrating over alternate variables,
we carry out the integration over the odd variables
$x_1,x_3,\ldots ,x_{2N-1}$ first and then over the remaining even variables.
The final result is

\begin{equation}
{\cal Z}_{2N} = K (2N)! \prod _{k=0}^{N-1} r_k         \label{11}
\end{equation}

So far the  problem is formally solved. In order to make the solution explicit,
the strategy consists of relating the set of skew orthogonal polynomials $R_i$
with the well-known set of orthogonal polynomials solving the Hermitian
one-matrix model. Thus, let us introduce orthogonal polynomials $C_i$ with
respect to the follwing symmetric scalar product

\[ <C_i,C_j>_C \equiv \int _{-\infty }^{\infty } dx
e^{-\beta V(x)} C_i(x) C_j(x)\]
\begin{equation}
\equiv  h_i \delta _{ij}                              \label{12}
\end{equation}

\n These polynomials are also monic and satisfy a two-step recursion relation

\begin{equation}
x C_i = C_{i+1} + \sigma _i C_i + \rho _i C_{i-1}        \label{13}
\end{equation}

The coefficient $\sigma _i$ takes into acount that we are dealing with a
non-even
potential (\ref{2}). In addition, it is possible to derive equations for
$\rho _i$, $\sigma _i$ from the reltions
$<C_i,C^{'}_i>_C=0$ and $<C_{i-1},C^{'}_i>_C=ih_{i-1}$. This yields to

\begin{equation}
<C_i,\beta V^{'} C_i>_C = 0                               \label{14}
\end{equation}

\begin{equation}
<C_{i-1},\beta V^{'} C_i>_C = i h_{i-1}                    \label{15}
\end{equation}

Now let us express the set of polynomials $R_i$ as a linear combination of the
set $C_i$.

\begin{equation}
R_i(x) = C_i(x) + a_{i,i-1} C_{i-1}(x) + a_{i,i-2} C_{i,i-2}(x) +
a_{i,i-3} C_{i,i-3}(x)                                           \label{16}
\end{equation}

In the next section we shall see that the combination (\ref{16}) ends at
$C_{i-3}$ because the order of the potential $V(x)$ is 3.

Neither $R_i$ nor $C_i$ have a definite parity anymore due to the cubic term
 in $V(x)$.  Then, it is no longer possible to split the relation (\ref{16})
into two, one for the set $R_{2i}$ and another for $R_{2i-1}$, as occurred
when dealing with even potentials \cite{Meh-M}, \cite{Bre-N}. This amounts to
an extra complication in solving (\ref{16}). At the end of the computation
of the coeffients $a_{i,j}$, we shall  make such a splitting to notice the
appearence of mixed terms between both sets.

To proceed further, we have to relate the two scalar products introduced in
(\ref{8}) and (\ref{12}). Given two functions $f$, $g$ it is easy to
prove, upon integration by parts, the following essential relation

\begin{equation}
<f,g>_C = -\frac{1}{2}<\beta V^{'} f,g>_R + <f^{'},g>_R     \label{17}
\end{equation}

\n This equation will give us enough relations to find out $a_{i,j}$, provided
we know the action of $V^{'}(x)$ and $\frac{d}{dx}$ on the set of orthogonal
polynomials $C_i$. In fact, computing $<C_{i-j},R_i>_C$, $j=0,1,2,3$ in
(\ref{16}) with the aid of (\ref{17}), it is possible to set up the
following relations for the unknown variables $a_{i,j}$

\[ a_{i,i-j} = <C_{i-j},R_i>_C \]
\begin{equation}
= -\frac{1}{2} <\beta V^{'} C_{i-j},R_i>_R +
<C^{'}_{i-j},R_i>_R \ \ \ \ j=0,1,2,3\  \forall i .           \label{18}
\end{equation}

\n For the action of $V^{'}(x)$ on $C_i$, using (\ref{2}) and (\ref{13}), we
have

\[ V^{'}(x) C_i(x) = g_3 C_{i+2} + [g_3(\sigma _{i+1}+\sigma _i) +
g_2]C_{i+1} \]
\[ + [g_3(\rho _{i+1}+\sigma ^2_i+\rho _i)+g_2 \sigma_i]C_i  \]
\[ + [g_3 \rho _i (\sigma _i+\sigma _{i-1})+g_2 \rho _i] C_{i-1} +
g_3 \rho _i \rho _{i-1}C_{i-2} \]
\begin{equation}
\equiv \sum _{l=-2}^{l=+2} v_{i,i+l} C_{i+l}(x)         \label{19}
\end{equation}

\n It is possible to simplify this expression using the discrete string
equations
(\ref{14}) and (\ref{15}). In fact, the following
expressions for the string equations are  readily obtained

\begin{equation}
\beta v_{i,i-1} = i =
 \beta [g_3 \rho _i (\sigma _i+\sigma _{i-1})+g_2\rho _i ]  \label{20}
\end{equation}

\begin{equation}
v_{i,i} = 0 = g_3 (\rho _{i+1}+\sigma _i^2+\rho _i) + g_2 \sigma _i \label{21}
\end{equation}

\n From (\ref{19}) and (\ref{20}) we can also express the coefficient
$v_{i,i+1}$
in a more tractable way

\begin{equation}
v_{i,i+1} = \frac{v_{i+1,i}}{\rho _{i+1}} =
 \frac{i+1}{\beta \rho _{i+1}}                              \label{22}
\end{equation}

\n On the other hand, the action of $d/dx$ on $C_i$ is given by

\begin{equation}
\frac{dC_i}{dx} = \sum _{j=0}^{i-1} \frac{<C_j,\beta V^{'}C_i>_C}
{<C_j,C_j>_C} \ C_j                                        \label{23}
\end{equation}

\n which is obtained by partial integration on $<C_j,\frac{dC_i}{dx}>_C$ with
$j<i$. As $V(x)$ has degree 3, this expression turns out simply to be

\begin{equation}
\frac{dC_i}{dx} =
 \beta v_{i,i-2} C_{i-2} + \beta v_{i,i-1} C_{i-1}           \label{24}
\end{equation}

\vspace{30pt}

{\bf 3.Recursion Relation for the Real Symmetric One-Matrix Model}

Due to the lengthy calculations involved for establishing and solving the
equations (\ref{18}), it is convenient to outline the main steps needed to
achieve the final result as follows,

i) Choose one of the four equations in (\ref{18}) ($j=0,1,2,3$).

ii) Express $V^{'}(x)C_{i-j}$ and $C^{'}_{i-j}$ in terms of the orthogonal
polynomials $C_i$. To do this, use (\ref{15})-(\ref{22}) and (\ref{24}).

iii) Express the $C_i$ polynomials of step ii) in terms of skew orthogonal
polynomials. To do this, it is required to invert the relations (\ref{16}).

iv) Use the skew-orthogonality relations (\ref{8}) and (\ref{9}) to solve
the right hand side of (\ref{18}).

Moreover, to carry out the above program we shall make repeatedly use of
the following essential observations,

a) As $R_i=C_i + \mbox{order ($C_{i-1}$)}$, then
$C_i=R_i + \mbox{order ($R_{i-1}$})$.

b) From the skew-orthogonality relations (\ref{8}),(\ref{9}) we have

\begin{equation}
<R_{2i},R_{2i+1}>_R \  = r_i = -<R_{2i+1},R_{2i}>_R             \label{25}
\end{equation}

and the rest of scalar products are zero.

c) The polynomial $R_{2i}$  is skew orthogonal to any polynomial of degree
less than or equal to $2i$.

The polynomial $R_{2i+1}$ is skew orthogonal to any polynomial of degree
less than $2i$. Simbolically,

\[ R_{2i}\ \  \mbox{skew} \bot \ \  R_i, R_{2i-1},\ldots \]
\begin{equation}
 R_{2i+1}\ \  \mbox{skew} \bot \ \ \ \ \ \ R_{2i-1},\ldots
\label{26}
\end{equation}

 To be consistent, we need to show that the relations (\ref{16})
actually stop at $C_{i-3}$. Namely, let us assume that a term, say,
$C_{i-4}$ is present. Then, the term in the r.h.s. of (\ref{18}) with
the highest $C_i$ is $<C_{i-4+2},R_i>_R$. This follows from (\ref{19})
and (\ref{23}). Now using observations a) and c) we easily get
$<C_{i-2},R_i>_R=0$.

We shall first work out the equations (\ref{18}) in the order $j=3,2,1$. The
expressions for $a_{i,i-3}$, $a_{i,i-2}$, $a_{i,i-1}$ thus obtained will be
 replaced in equation (\ref{18}) for $j=0$. This will turn out to be the
desired recursion relation that solves the discrete model.

{\em Equation (\ref{18}) \mbox{$j=3$}}.

\begin{equation}
a_{i,i-3} h_{i-3} = -\frac{\beta }{2} v_{i-3,i-1} <R_{i-1},R_i>_R \label{27}
\end{equation}

It is convenient to split the indices into even and odd.
If $i=2k$, then as $<R_{2k-1},R_{2k}>_R=0$, one deduces

\begin{equation}
a_{2k,2k-3} = 0                                                \label{28}
\end{equation}

If $i=2k+1$, as $<R_{2k},R_{2k+1}>_R=r_k$ and $v_{i-3,i-1}=g_3$, one deduces

\begin{equation}
a_{2k+1,2k-2} h_{2k-2} = -\frac{\beta }{2} g_3 r_k              \label{29}
\end{equation}

{\em Equation (\ref{18}) \mbox{$j=2$}}.

It is convenient to start with

\begin{equation}
a_{i,i-2} h_{i-2} = -\frac{\beta }{2} v_{i-2,i} <C_{i},R_i>_R
-\frac{\beta }{2} v_{i-2,i-1} <C_{i-1},R_i>_R                \label{30}
\end{equation}

After some algebra we get

\begin{equation}
a_{i,i-2} h_{i-2} = -\frac{\beta }{2}[v_{i-2,i}a_{i,i-1}+v_{i-2,i-1}]
<R_{i-1},R_i>_R                                             \label{31}
\end{equation}

If $i=2k$ then,

\begin{equation}
a_{2k,2k-2} = 0                                               \label{32}
\end{equation}

If $i=2k+1$ then,

\begin{equation}
a_{2k+1,2k-1} h_{2k-1} = -\frac{\beta }{2}[g_3 a_{2k+1,2k} + \frac{2k}
{\beta \rho _{2k}}]\ r_k                                        \label{33}
\end{equation}

and $a_{2k+1,2k}$ is to be determined.

{\em Equation (\ref{18}) \mbox{$j=1$}}.

Applying again the procedure previously described, we obtain

\[
\frac{2}{\beta } a_{i,i-1} h_{i-1} = -v_{i-1,i+1}<R_{i+1},R_i>_R
\]
\begin{equation}
 - [v_{i-1,i+1}(a_{i+1,i} a_{i,i-1}-a_{i+1,i-1})+v_{i-1} a_{i,i-1}]
<R_{i-1},R_i>_R                                               \label{34}
\end{equation}

\n Where we have kept in mind that $v_{i-1,i-1}=0$ due to the string equation
(\ref{21}). Notice that in (\ref{34}) appears $a_{i+1,i-1}$, which is of the
same
type described by the previous case (\ref{31}). Replacing the value of
$a_{i+1,i-1}$ in (\ref{34}) we shall obtain an equation for $a_{i,i-1}$ solely.
However, the analysis is simplifyed if we first make the splitting of indices
into even and odd.

If $i=2k$, then $<R_{i-1},R_i>_R$ = 0 and

\begin{equation}
\frac{2}{\beta } a_{2k,2k-1} h_{2k-1} = g_3 r_k                \label{35}
\end{equation}

If $i=2k+1$, then $<R_{i+1},R_i>_R = 0$ and

\begin{equation}
\frac{2}{\beta } a_{2k+1,2k} h_{2k} = -g_3 a_{2k+1,2k }[a_{2k+2,2k+1}+
\frac{2k+1}{\beta \rho _{2k+1}}] r_k                            \label{36}
\end{equation}

\n This equation admits two solutions. We shall take the simplest one,

\begin{equation}
a_{2k+1,2k} = 0                                              \label{37}
\end{equation}

\n The case $a_{2k+1,2k}\neq 0$ is considered in Appendix I. Now we can replace
 (\ref{37}) into (\ref{33}) to obtain the value of $a_{2k+1,2k-1}$,

\begin{equation}
a_{2k+1,2k-1} = -\frac{k}{\rho _{2k}}\ \frac{r_k}{h_{2k-1}}    \label{38}
\end{equation}

Now we have the solution of all the coefficients relating both sets of
polynomials in (\ref{16}). With this solution, it is convenient to split
the relation (\ref{16}) into polynomials with even and odd indices to
see the picture that emerges from the previous calculations. Therefore,

\begin{equation}
R_{2k} = C_{2k} + a_{2k,2k-1} C_{2k-1}                        \label{39}
\end{equation}

\begin{equation}
R_{2k+1} = C_{2k+1} + a_{2k+1,2k-1} C_{2k-1} +
 a_{2k+1,2k-2} C_{2k-2}                                        \label{40}
\end{equation}

\n where the coefficients are given by (\ref{35}), (\ref{37}) and (\ref{40}).
In this way, we observe that there are odd contributions to $R_{2k}$ as
well as even ones to $R_{2k+1}$. This is the new effect of dealing with
a non-even potential, causing the polynomials $R_{2k}$, $R_{2k+1}$ not
to have a definite parity anymore.

 Now we are in position to set up the recursion relation between the
relevant quantities $r_k$, $h_k$ that ultimately solves our model. This is
 achieved with equation (\ref{18}) for $j=0$. The main steps of this lengthy
calculation
 is given in Appendix II with the following result,

\begin{equation}
 h_{2k} = [-\frac{\beta ^2}{4}\ g_3^2\ \frac{r_{k+1}}{h_{2k+1}} +
\frac{2k+1}{2\rho _{2k+1}}]\ r_k                              \label{41}
\end{equation}

The equation (\ref{18}) for $j=0$ are actually two relations, depending on
whether the index $i$ is even or odd, as was done in the previous cases. This
means that the system of equations is constrained, for there is one more
equation that unknown variables. It is worth noticing that either equation
(\ref{18}) for $j=0$ leads to the same solution (see Appendix II), thereby
providing a proof of the consistency of our solution. In addition, if we set
$g_3=0$ we should end up
with the solution of the gaussian model \cite{Meh-M}, \cite{Bre-N}, as it
happens

\begin{equation}
h_{2k} = \frac{\beta }{2}\ r_k                               \label{42}
\end{equation}

\n for $\rho _i=i/\beta $ in the gaussian case.

{\bf 4.The Partition Function of the Model}

So far, we have related the calculation of the partition function for
the real symmetric one-matrix model to the solution of the Hermitian
ensemble. Once the norms of the orthogonal polynomials $C_i$ are determined
from the discrete string equations (\ref{20}), (\ref{21}), then the
quantities $r_k$ are in principle computed from (\ref{41}) yielding the
solution of the partition function (\ref{11}). In fact, the recursion
relation(\ref{41}) obtained for $r_k$ amounts to a recursion relation
between partition functions at different $N$. From (\ref{11}) it is
possible to rewrite $r_k$ in the form

\begin{equation}
r_k = \frac{1}{(2k+2)(2k+1)}\times
\frac{\mbox{{\cal Z}}_{2k+2}}{2k}                       \label{43}
\end{equation}

Upon substitution of (\ref{43}) in (\ref{41}) we readily obtain the
following recursion relation for the partition function,

\[ \frac{\beta ^2}{4}g_3^2 {\cal Z}_{2k+4} - \frac{1}{2}(2k+4)(2k+3)(2k+1)
h_{2k}{\cal Z}_{2k+2}  \]
\begin{equation}
+ \frac{(2k+4)!}{(2k)!}h_{2k+1}h_{2k}
{\cal Z}_{2k} = 0                                        \label{44}
\end{equation}

When the continuum limit of the model is to be taken, we shall see that it is
convenient to introduce the following quantities relted to the ``norms"
$r_k$ by

\begin{equation}
W_k = \beta \frac{r_k}{h_{2k}}                           \label{45}
\end{equation}

\n and the partition function (\ref{11}) is determined by the quantities
$W_k$ and $h_{2k}$ as

\begin{equation}
{\cal Z}_{2N} = K\times (2N)!\beta ^{-(N-1)}
\prod _{k=0}^{N-1} W_k h_{2k}                             \label{46}
\end{equation}

\n Hence, the free energy takes the form

\begin{equation}
{\cal F}_{2N} = \ln K_N + \sum _{k=0}^{N-1}\ln h_{2k} +
\sum _{k=0}^{N-1} W_{2k}                                   \label{47}
\end{equation}

According to \cite{Bre-N},\cite{Har-M}, the second term in ${\cal F}_{2N}$
represents the contribution of the orientable surfaces to the free energy,
for  it enterely involves the constants related to the Hermitian matrix
model which describes only oriented surfaces. On the other hand, the
constants $W_{k}$ amounts to the contribution of non-orientable surfaces
to the free energy. This is made more plausible by noticing that the
odd powers of ($1/N$) in ${\cal F}_{2N}$ come precisely from $W_k$
\cite{Bre-N}.

{\bf 5.The Continuum Limit}

We now want to take the continuum limit of our solution (\ref{41}) to
describe the critical point of non-orientable random surfaces so, we
need to borrow the results concerning the double-scaling limit for the
Hermitian model with a cubic interaction.

The scaling limit of the string equations (\ref{20}),(\ref{21}) goes as
follows. First, we take the planar limit ($N\rightarrow \infty $, with
$\frac{N}{\beta }\rightarrow 1$) of (\ref{20}),(\ref{21}) at $i=2N$.
Therefore, it is convenient to introduce the functions

\begin{equation}
V_1(\rho ,\sigma ) = 2\rho + g_3 \rho \sigma = 1              \label{48}
\end{equation}

\begin{equation}
V_2(\rho ,\sigma ) = 2\sigma + g_3 (2\rho + \sigma ^2) = 0    \label{49}
\end{equation}

\n where we have normalized $g_2=4$ without loss of generality. Then, the
critical values of $g_{3c}$ and $\rho _c$ are defined by

\begin{equation}
\left. \frac{d V_1(\rho ,\sigma )}{d\rho }\right| _c = 0      \label{50}
\end{equation}

\n subject to

\begin{equation}
\left. \frac{d V_2(\rho ,\sigma )}{d\rho }\right| _c = 0       \label{51}
\end{equation}

The analysis of these conditions leads to the following critical values
that, for future convenience, we recast as

\begin{equation}
\rho _c^2 = \frac{3}{4} \ \ \ \mbox{and} \ \
g_{3c}^2\rho _c = \frac{4}{3}                                 \label{52}
\end{equation}

Second, we insert the values (\ref{52}) in the discrete string equations
(\ref{20}),(\ref{21}) at $i=2N$ and consider the following scaling
ansatze \cite{My-P1} \cite{Bre-N} \cite{Sim}\footnote{Let us notice that
this is the same ansatze as is \cite{Sim} but in a different notation.}

\begin{equation}
\rho _{2N+k} = \rho _c [1 - \delta ^2 f(t_k)]              \label{53}
\end{equation}

\begin{equation}
\sigma _{2N+k} =  - \delta ^2 s(t_k)                     \label{54}
\end{equation}

where

\begin{equation}
\delta = \beta ^{-1/5} \rightarrow 0                       \label{55}
\end{equation}

\begin{equation}
t_k = (2\beta -2N-k) \delta                               \label{55b}
\end{equation}

The doble-scaling limit is achieved when $\delta \rightarrow 0$ and
$\frac{\beta }{N} \rightarrow 1$ while $t=t_0$ is held fixed. Following
reference \cite{Sim}\footnote{I am very grateful to S.Dalley for explaining
this point to me and drawing my attention to ref.\cite{Sim} during the course
of this work.}, it is
appropiate to introduce two auxiliary scaling functions $f_{\pm }(t_k)$ defined
in terms of $\rho (t_k)$ and $\sigma (t_k)$ by

\begin{equation}
f_{\pm}(t_k) = f(t_k) \pm s(t_k)                             \label{56}
\end{equation}

When dealing with non-even potentials, the key point to obtain the differential
equations for $f_{+}$ and $f_{-}$ in the limit $\delta \rightarrow 0$ is to
express $f(t_k)$ and $s(t_k)$ again as a perturbative series in the $\delta $
parameter\footnote{We refer the reader to ref.\cite{Sim} for details.}

\begin{equation}
f(t) = f_0(t) + f_1(t)\delta + f_2(t)\delta ^2 + \cdots        \label{57}
\end{equation}

\n with

\begin{equation}
f_0(t) = \frac{1}{2}\rho _{-}(t)                             \label{58}
\end{equation}

\n and

\begin{equation}
s(t) = s_0(t) + s_1(t)\delta + s_2(t)\delta ^2 + \cdots      \label{59}
\end{equation}

As far as the leading behaviour of the partition function is concerned, we
only need to know about the differential equations satisfied the first
terms in the expressions (\ref{57}), (\ref{59}). These are found to satisfy
\cite{Sim} \cite{Vis} (up to trivial rescaling) the Painle\'e I equation

\begin{equation}
f_0^2 - \frac{1}{3} f_0^{''} = t                               \label{60}
\end{equation}

\n suplemented by the equation

\begin{equation}
\left. \rho _+\right|_{\delta =0} = f_0 + s_0 = 0                \label{61}
\end{equation}

This completes the analysis of the Hermitian model with a cubic potential. To
begin with the continuum limit of the real symmetric model, we rewrite the
recursion relation at $i=N$ in the following form

\begin{equation}
\frac{g_3^2}{4}\ W_{N+1}W_{N}\ \rho _{2N+2}\rho _{2N+1}
-\frac{2N+1}{2\beta }\ W_N + \rho _{2N+1} = 0               \label{62}
\end{equation}

\n where $W_N$ is given by (\ref{45}). This is a quadratic equation in the
unknown variable $W$. The spherical limit of (\ref{62}) is achieved by
taking $N,\beta \rightarrow \infty $ with $N/\beta $ finite and
assuming $W_N\rightarrow W_c$, $\rho _N \rightarrow \rho _c$. Tuning
$g_3$ and $\rho $ to their critical values (\ref{52}) previously found,
we observe that (\ref{62}) has a unique root $W_c$ given by

\begin{equation}
W_c = 2\rho _c                                              \label{63}
\end{equation}

Let us notice though that $\rho _c$ can take two possible values according
to (\ref{52}). So, the critical behaviour  of the Hermitian case leads also
to critical behaviour in the symmetric model.

The equation (\ref{62}) is similar to the ones appearing in the analysis of
the real symmetric ensemble with a cuartic potential \cite{Bre-N}, \cite{Har-M}
 and in the simplectic ensemble \cite{My-P1}. The technical difference is that
our equation is quadratic in the unknown variable $W$ while those in the
forementioned references are cubic. Therefore, we shall use the same scaling
ansatze for
$W$ as in \cite{My-P1}, namely

\begin{equation}
W_{N+k} = W_c\ e^{\delta \omega (t_{2k})}                          \label{64}
\end{equation}

\n where $\delta = \beta ^{-1/5}$, and $W_c$, $t_k$ are given by (\ref{63}),
(\ref{55b}) respectively. The novelty of working with a non-even potential
is that $\omega (t_k)$, with $t\equiv t_0$, has to be expanded in a series of
 $\delta $

\begin{equation}
\omega (t) = \omega _0(t) + \omega _1(t) \delta
+ \omega _2(t) \delta ^2 + \cdots                                 \label{65}
\end{equation}

\n in agreement with the procedure developed in (\ref{57}) for  the functions
$\rho _N$. As far as  the leading behaviour of the partition function is
concerned, we are only interested in the differential equation obeyed by the
term
$\omega _0$ in (\ref{65}). The way to proceed is to insert the ansatze
(\ref{53}) and (\ref{64}) in the recursion relation (\ref{62}) and to expand in
$\delta $ with the aid of (\ref{55b}). Then, we solve for the $\omega 's$ in
$\ref{65}$ order by order  until we obtain a differential equation that
determines
$\omega _0$. This turns out to be,

\begin{equation}
3f_0 = \omega _0^2  - 2\omega _0^{'}                             \label{66}
\end{equation}

We can thought of (\ref{66}) as an inhomogeneous ordinary differential equation
yielding the solution for the unorientable contribution $\omega _0$ to the free
energy once the orientable contribution $f_0$ is known after solving the
Painlev\`e I equation (\ref{60}). This is the same kind of differential
equation\footnote{Up to trivial rescalings of $f_0$, $\omega _o$ and $t$ as in
(\ref{60}).} that appears in the analysis of the real symmetric ensemble
carried out with a cuartic potential \cite{Bre-N}, \cite{Har-M}. To be precise,
the
general solution found out with a cuartic interaction also includes in
(\ref{66}) a term of the type $c e^{\int _1^t \omega _0 (s) ds}$, where $c$ is
a constant of integration. Our solution corresponds to the choice $c=0$. This
$c$-term corresponds to different non-perturbative contributions to $\omega _0$
but, as far
as the finite genus contributions is concerned, there is no difference of
information among the possible choices of $c$. Thus, the  asymptotics series
of $\omega _0$ is $c$-independent \cite{Bre-N} and is given by

\begin{equation}
\omega _0 = \sum _0^{\infty } \omega _n^{(0)}
 t^{-(5n-1)/4}                                         \label{67}
\end{equation}

\n where now there are even and odd contributions to the free energy as an
expansion of integer powers of $1/N$. It is also straighforward to rewrite the
expression of the free energy $\phi $ \cite{Bre-N}, \cite{Har-M} in the
presence of a cubic interaction,

\begin{equation}
 \phi = f_0 + \omega ^{'}_0                            \label{68}
\end{equation}

 Let us finally remark that when $c=0$, the mapping between $\omega _0$ and
$f_0$ turns out to be an ordinary Miura transformation \cite{Bre-N} \cite{Bel}.

{\bf 6.Concluding Remarks}

There is a good understanding of the features exhibited by the Hermitian
ensemble in its several versions such as the one-matrix model, multimatrix
extensions \cite{I-Z} \cite{Meh2}
           and the Hermitian matrix model in the presence of an external
field (the Kontsevich model) \cite{G-N1}, \cite{G-N2}, \cite{Kon}. However,
if we exchange the Hermitian ensemble with the real symmetric ensemble in
the forementioned models, our knowledge decreases drastically. Even in the
simplest case of a real symmetric one-matrix model, our understanding is not
that good. This makes it quite interesting to check the properties of the
real symmetric matrices in comparison with the equivalent properties for the
Hermitian matrices, as has been done in this letter for the case of
 pure gravity.

Following this philosophy, it would be quite interesting to know whether the
partition function of the real symmetric model is the $\tau $-function
of a certain hierarchy of integrable differential equations. The Miura
transformation (\ref{66}) might be helpful for this purpose. Unfortunately,
this is not the same Miura transformation relating the KdV hierarchy
associated with       $(\partial ^2 + f_0)$ to
its partner known as the mKdV hierarchy \cite{Bre-N}.

There is a natural extension of the present work to check whether or not
there exits a doubling of the ordinary differential equation (\ref{66})
when a more general non-even potential is considered \cite{In}. The simplest
case
in which this feature should appear is with a potential like
$V(x)=g_2 x^2+g_3 x^3+g_4 x^4$, as it happens in the Hermitian model
\cite{Bac-P} \cite{Sim}.

{\bf  Acknowledgements}

I would like to thank S.Dalley, U.Danielsson, D.Gross, I.Klebanov, \linebreak
M.Newman,
A.Pasquinucci and H.Verlinde for useful discussions and
explanations.
I also thank D.Katz and
E.Stark for their comments on the manuscript.

I wish to thank Ministerio de Educacion y Ciencia (Espa\~{n}a) for financial
support through a BFPIE postdoctoral fellowship.

{\bf Appendix I}

Let us analyze the other possibility for the equation (\ref{36}). If
$a_{2k+1,2k}\neq 0$, we end up with the following solution for the
element $a_{2k+2,2k+1}$,

\begin{equation}
a_{2k+2,2k+1} = -\frac{2}{\beta g_3} \frac{h_{2k}}{r_k}
-\frac{2k+1}{\beta \rho _{2k+1}}                            \label{I1}
\end{equation}

But let us notice that this is precisely an element of the type $a_{2k,2k-1}$
already determined in equation (\ref{35}). Thus, comparing the equations
(\ref{I1}) and (\ref{35}) we obtain now a recursion relation between $r_k$
and $h_{2k}$

\begin{equation}
h_{2k} = -\frac{g_3}{2}\ [ \frac{\beta ^2 g_3}{2} \frac{r_{k+1}}{h_{2k+1}}
+ \frac{2k+1}{\rho _{2k+1}}]\ r_k                          \label{I2}
\end{equation}
However, $a_{2k+1,2k}$ have not been determined yet. To do this, we have to
use the remainig equation (\ref{18}) for $j=0$. In this way, the roles of
equations (\ref{18}) $j=1$ and $j=0$ have been exchanged.

It is convenient again to split indices in (\ref{18}) $j=0$ in evens and odds.
If $i=2k$, as $<R_{2k-1},R_{2k}>_R=0$ and $<R_{2k+1},R_{2k}>=-r_k$, we have

\begin{equation}
h_{2k} = \frac{\beta }{2}[ g_3 a_{2k+2,2k+1} -
\frac{2k+1}{\beta \rho _{2k+1}}]  (-r_k)                  \label{I3}
\end{equation}

\n and this determines the last coefficient that we were left with being

\begin{equation}
a_{2k+2,2k+1} = -\frac{2}{g_3 \beta } \frac{h_{2k}}{r_k} +
\frac{2k+1}{g_3 \beta \rho _{2k+1}}                        \label{I4}
\end{equation}

Now, comparing the equations (\ref{I1}) an (\ref{I4}) we draw the conclusion
that the possibility we are dealing with is only consistent iff the coupling
constant takes the value

\begin{equation}
g_3 = -1                                                      \label{I5}
\end{equation}

In addition we have still to impose the relation coming from equation
(\ref{18}) $j=0$ when the indices are odd. Unlike the case of the simple
solution (\ref{37}), now this relation takes an extremely long and cumbersome
form
due to the fact that most of the $a$-coefficients do not vanish.
Nevertheless, as the coupling constant is absolutely fixed by (\ref{I5}), the
possibility (\ref{I1}), if consistent, would not be relevant as far as the
contunuum limit is concerned.

{\bf Appendix II}

To obtain the recursion relation (\ref{41}) solving the model, we start with
equation (\ref{18}) for $j=0$. After some algebra involving the use of  the
observations described in section 3, we arrive at

\[ h_i = -\frac{\beta }{2}v_{i,i+2}<C_{i+2},R_i>_R \]
\begin{equation}
-\frac{\beta }{2}v_{i,i+1}<C_{i+1},R_i>_R
-\frac{\beta }{2}v_{i,i-1}<C_{i-1},R_i>_R                 \label{II1}
\end{equation}

\n Using (\ref{16}) to invert the relation between $R_i$ and $C_i$, we find

 \[ h_i = \frac{\beta }{2}[v_{i,i+2}a_{i+2,i+1}-v_{i,i+1}]
<R_{i+1},R_i>_R \]
\[ -\frac{\beta }{2} \{ v_{i,i+2}[-(a_{i+2,i+1} a_{i+1,i}-a_{i+2,i}) a_{i,i-1}
+(a_{i+2,i+1} a_{i+1,i-1}-a_{i+2,i-1})] \]
\begin{equation}
+v_{i,i+1}(a_{i+1,i} a_{i,i-1}-a_{i+1,i-1}) - v_{i,i-1} \}
 <R_{i-1},R_i>_R                                            \label{II2}
\end{equation}

\n It is useful again to split the indices into even and odd. If $i=2k$, as
$<R_{2k-1},R_{2k}>_R=0$
and
$<R_{2k+1},R_{2k}>_R=-r_k$, it follows

\begin{equation}
a_{2k+2,2k+1} = \frac{\beta }{2} g_3 \frac{r_{k+1}}{h_{2k+1}} \label{II3}
\end{equation}

\n Replacing the value of $a_{2k+2,2k+1}$ from (\ref{35}) we obtain the desired
result (\ref{41}) for the recursion relation,

\begin{equation}
h_{2k} = [-\frac{\beta ^2}{4} g_3^2 \frac{r_{k+1}}{h_{2k+1}} +
\frac{2k+1}{2\rho _{2k+1}}] r_k                            \label{II4}
\end{equation}

If $i=2k+1$, the analysis in principle is more cumbersome. Now we have
$<R_{2k+2},R_{2k+1}>_R=0$
and
$<R_{2k},R_{2k+1}>_R=r_k$. Fortunately, using the equations (\ref{28}),
(\ref{32}) and (\ref{37}) it easy to see that all the constants $a's$
entering the equation (\ref{II2}) are now zero, except for
$a_{i+2,i-1}=a_{2k+3,2k}=-\frac{\beta }{2}g_3 \frac{r_{k+1}}{h_{2k}}$.
Then, after using (\ref{19}), the equation takes the form

\begin{equation}
h_{2k+1} = [-\frac{\beta ^2}{4} g_3^2 \frac{r_{k+1}}{h_{2k}} +
\frac{2k+1}{2}] r_k                                     \label{II5}
\end{equation}

\n Dividing (\ref{II5}) by $\rho _{2k+1}=\frac{h_{2k+1}}{h_{2k}}$ we achieve
precisely the equation (\ref{II4}) again.


\begin{thebibliography}{99}


\bibitem{B-K} E. Br\'ezin and V. Kazakov,
{\em Phys.\ Lett.\ \/}{\bf 236 B}
(1990) 144.

\bibitem{D-S} M.Douglas and S.Shenker,
{\em Nucl.\ Phys.\ \/}{\bf B 335}
(1990) 635 .

\bibitem{G-M} D. Gross and A. Migdal,
{\em Phys.\ Rev. \ Lett.\ \/}{\bf 64}
(1990) 127.

{\em Nucl.\ Phys.\ \/}{\bf B 340}
(1990) 333 .

\bibitem{Per-S} V. Periwal and D. Shevitz,
{\em Phys.\ Rev. \ Lett.\ \/}{\bf 64}
(1990) 1326.

\bibitem{My-P1} R.C. Myers and V. Periwal,
{\em Phys.\ Rev. \ Lett.\ \/}{\bf 64}
(1990) 3111.

\bibitem{My-P2} R.C. Myers and V. Periwal,
{\em Phys.\ Rev. \ Lett.\ \/}{\bf 65}
(1990) 1088.

\bibitem{Bre-N} E. Br\'ezin and H. Neuberger,
{\em Nucl.\ Phys.\ \/}{\bf B350 }
(1991)  513.

\bibitem{Bre-N2} E. Br\'ezin and H. Neuberger,
{\em Phys.\ Rev. \ Lett.\ \/}{\bf 65} (1990) 2098.

\bibitem{Har-M} G.R. Harris and E.J. Martinec,
{\em Phys.\ Lett.\ \/}{\bf  B245}
(1990) 384.

\bibitem{Meh-M} M.L. Mehta and G. Mahoux,
Preprint Saclay SPht/90-084.

\bibitem{Bac-P} C. Bachas and P.M.S. Petropoulos,
{\em Phys.\ Lett.\ \/}{\bf  B247}
(1990) 363.


\bibitem{Sim} S. Dalley, C.V. Johnson and T. Morris,
{\em \ Mod. \ Phys.\ Lett.\ \/}{\bf A6}
(1991) 439.

\bibitem{Vis} C. Marzban and R.R. Viswanathan,
{\em \ Int.\ J. \ Mod. \ Phys.\ \/}{\bf  A6}
(1991) 2559.


\bibitem{Meh1} M.L. Mehta,
{\em Random Matrices \ \/}
(Academic Press, New York, 1967)

\bibitem{Bel}A.A. Belavin {\em KdV equations and W-algebras \ \/}
 (Lecture given at Taniguchi Foundation Symposium, Integrable Models in Quantum
Field Theory and Statistical Physics, Kyoto 1988).


\bibitem{I-Z} C. Itzykson and J.-B. Zuber,
{\em J.\  Math.\ Phys.\
\/}{\bf 21} (1980) 411.


\bibitem{Meh2} M.L. Mehta,
 {\em Comm.\  Math.\ Phys.\
\/}{\bf 79} (1981) 327.


\bibitem{G-N1} D. Gross and M. Newman,
{\em Phys.\ Lett.\ \/}{\bf  B266}
(1991) 291.

\bibitem{G-N2} D. Gross and M. Newman,
Princeton Preprint PUPT-1282 Dec. 1991


\bibitem{Kon} M. Kontsevich,
{\em Funk. \ An. \ Appl.\ \/}{\bf  25}
(1991) 50.


\bibitem{In} Work in progress.










\end{thebibliography}
\end{document}